\documentclass[preprint,12pt]{elsarticle}

\usepackage{booktabs}
\usepackage{bm}
\usepackage{amsmath,amssymb,amsfonts}
\usepackage{algorithmic}
\usepackage{graphicx}
\usepackage{rotating}

\usepackage{textcomp}
\usepackage{color}

\begin{document}

\begin{frontmatter}



\title{Leveraging Large Self-Supervised Time-Series Models for Transferable Diagnosis in Cross-Aircraft Type Bleed Air System}


\begin{abstract}
Bleed Air System (BAS) is critical for maintaining flight safety and operational efficiency, supporting functions such as cabin pressurization, air conditioning, and engine anti-icing. However, BAS malfunctions, including overpressure, low pressure, and overheating, pose significant risks such as cabin depressurization, equipment failure, or engine damage. Current diagnostic approaches face notable limitations when applied across different aircraft types, particularly for newer models that lack sufficient operational data. To address these challenges, this paper presents a self-supervised learning-based foundation model that enables the transfer of diagnostic knowledge from mature aircraft (e.g., A320, A330) to newer ones (e.g., C919). Leveraging self-supervised pretraining, the model learns universal feature representations from flight signals without requiring labeled data, making it effective in data-scarce scenarios. This model enhances both anomaly detection and baseline signal prediction, thereby improving system reliability. The paper introduces a cross-model dataset, a self-supervised learning framework for BAS diagnostics, and a novel Joint Baseline and Anomaly Detection Loss Function tailored to real-world flight data. These innovations facilitate efficient transfer of diagnostic knowledge across aircraft types, ensuring robust support for early operational stages of new models. Additionally, the paper explores the relationship between model capacity and transferability, providing a foundation for future research on large-scale flight signal models.
\end{abstract}

\author[mymainaddress,second]{Yilin Wang\fnref{equal}}

\author[mymainaddress]{Peixuan Lei\fnref{equal}}
\author[mymainaddress,third]{Xuyang Wang}
\author[third]{Liangliang Jiang}
\author[third]{Liming Xuan}
\author[fourth]{Wei Cheng}
\author[fourth]{Honghua Zhao}
\author[mymainaddress]{Yuanxiang Li\corref{mycorrespondingauthor}}
\cortext[mycorrespondingauthor]{Corresponding author: School of Aeronautics and Astronautics, Shanghai Jiao Tong University, Shanghai 200240, China. Tel: +86 21-34206160.}
\fntext[equal]{These authors contributed equally to this work.}
\address[mymainaddress]{School of Aeronautics and Astronautics, Shanghai Jiao Tong University, Shanghai, China}
\address[second]{Shanghai Innovation Institute, Shanghai, China}
\address[third]{Shanghai Aircraft design and research Institute, Shanghai, China}
\address[fourth]{China Eastern Airlines Technic Company Co.Ltd, Shanghai, China}

\begin{keyword}


Fault Diagnosis, Anomaly Detection,Self-supervised Learning, Foundation Models, Civil Aircraft
\end{keyword}

\end{frontmatter}



\section{Introduction}
\label{sec1}

Bleed Air System (BAS) is crucial for ensuring flight safety and operational efficiency, managing the air supply for systems such as cabin pressurization, air conditioning, and engine anti-icing \cite{adib2007aircraft,hodal2005bleed,michaelis2018aircraft}. As a key component of an aircraft's pneumatic system, the BAS extracts air from the engine compressor and distributes it to these essential subsystems. Despite their importance, BAS are prone to malfunctions, including overpressure, low pressure, and overheating. These faults can lead to severe safety risks, such as cabin depressurization, equipment failure, or engine damage if left unresolved \cite{rios2013mobile}. For instance, overpressure can damage sensitive components, low pressure may compromise the air supply, affecting cabin environment control, and overheating could result in system failures or fire hazards \cite{maitre2016phm}. Given their frequent occurrence and potential impact, timely and effective diagnosis of BAS malfunctions is essential to avoid catastrophic outcomes.

Several diagnostic and predictive methods have been proposed to tackle general fault diagnosis challenges across industrial systems \cite{lin2024channel,zio2022prognostics,zhang2022prediction,wang2023self,jiang2024attention,wang2024phyformer}. These approaches include model-based\cite{venkatasubramanian2003review,ma2022data}, data-driven, and hybrid methods \cite{venkatasubramanian2003review,solomatine2008data,schwabacher2005survey}. Model-based techniques leverage physical models to simulate system behavior and detect deviations, while data-driven methods rely on historical data to uncover patterns and predict failures\cite{zhang2016data,wen2017new,schwabacher2005survey,wu2024data,wang2024model}. Hybrid methods combine the strengths of both approaches. Recent studies have further improved diagnostic accuracy by addressing challenges such as imbalanced data distribution, domain adaptation, and abnormal signal recovery. Advances in adaptive clustering weighted oversampling \cite{LI2024RESS} and imbalance-aware predictive maintenance techniques \cite{Li2023Sensors} have enhanced fault diagnosis under data-limited conditions. Additionally, multi-sensor data fusion techniques \cite{JIANG2024Multi} and contrast-assisted domain adaptation methods \cite{Jiang2025Contrast} have been introduced to improve robustness in varying operational environments, enabling more effective generalization across different system conditions. Furthermore, the variable scale multilayer perceptron \cite{FAN2024EAAI} has been proposed to recover abnormal vibration signals in complex mechanical systems, such as helicopter transmission systems, providing a more efficient and reliable solution for predictive maintenance. However, these methods have not been specifically tailored to the unique requirements of BAS diagnostics, which involve highly variable operational conditions and complex system dependencies. Moreover, their reliance on extensive labeled datasets limits their effectiveness for new aircraft types, where operational data is often sparse\cite{hu2021task}.

Traditional research on the Bleed Air System (BAS) has primarily focused on data-driven health monitoring and risk assessment using statistical and machine learning methods. Existing approaches include dynamic health index construction \cite{wang2024data}, multivariate state estimation techniques for risk warning \cite{su2022risk}, and multi-level feature extraction for health indicator construction \cite{duan2023method}. While these studies have demonstrated the feasibility of BAS fault diagnosis, they typically rely on aircraft-specific data distributions, limiting their adaptability to different aircraft models. Moreover, their dependence on extensive labeled datasets restricts their effectiveness in early operational stages, particularly for newly deployed aircraft such as the C919, where operational data is scarce.

To address these limitations, there is a growing need for universal models capable of transferring diagnostic knowledge across different aircraft types. Such models can reduce the reliance on large datasets, enabling diagnostics based on shared system characteristics. However, developing unified models faces several challenges, including variations in signal characteristics, fault causes, and operational conditions across aircraft types. Furthermore, the fragmented and often incomplete nature of diagnostic data complicates model development, making it difficult to establish a robust cross-aircraft diagnostic framework.

This paper proposes a self-supervised learning-based foundation model to address the challenges of diagnosing BAS issues across different aircraft types. By leveraging self-supervised pretraining\cite{zou2022spot,ericsson2021well,bao2024simtseg}, the model extracts feature representations from flight signals without the need for labeled data, allowing for the transfer of diagnostic knowledge from mature aircraft types (e.g., Airbus A320\cite{gerdes2017genetic},Airbus A330) to newer models (e.g., C919). This approach is particularly valuable during the early operational phases of new aircraft, where data is often limited. The model excels in both baseline signal prediction and anomaly detection, providing interpretable diagnostic results while reducing reliance on large labeled datasets. By enhancing adaptability across various aircraft types, the model ensures that new aircraft receive crucial operational support during their early deployment stages.

This paper presents three key innovations in the field of transferable fault diagnosis for aircraft Bleed Air Systems (BAS):

(1) \textbf{A large-scale cross-aircraft dataset:} We construct a comprehensive dataset covering multiple aircraft types (A320, A330, and C919), facilitating research on cross-model diagnostic transferability. This dataset provides a foundation for studying fault diagnosis across different aircraft platforms, addressing the challenge of data scarcity in newly deployed models.

(2) \textbf{A scalable self-supervised learning framework for BAS diagnostics:} We develop a novel self-supervised learning approach that enables the model to learn transferable feature representations from flight signals without requiring extensive labeled fault data. Furthermore, we investigate the relationship between model scale and diagnostic performance, demonstrating that larger models exhibit superior transferability across aircraft types. This insight highlights the potential of scaling up self-supervised models to improve generalization capabilities in fault diagnosis, paving the way for future research on large-scale flight signal models.

(3) \textbf{A unified diagnostic loss function and knowledge transfer mechanism:} We introduce a Joint Baseline and Anomaly Detection Loss Function that integrates baseline signal prediction with anomaly detection, allowing for efficient fine-tuning and improved interpretability. Additionally, we demonstrate the effective transfer of diagnostic knowledge from mature aircraft to newer models, ensuring reliable early-stage operational support.

These contributions collectively establish a scalable and transferable fault diagnosis system, advancing the state-of-the-art in aviation diagnostics. Our work tackles critical challenges related to data scarcity and model generalization while also laying the foundation for future research on large-scale flight signal models and their broader applications in aircraft maintenance and safety.


\section{Problem Definition}
\label{sec2}
\subsection{Cross-Aircraft Type Modeling Challenges}
\label{subsec1}
The diagnosis of BAS for different types of aircraft presents significant challenges due to the inherent differences in each system. During operation, each aircraft type generates distinct datasets $D=\{D_i \mid i=1, \ldots, n\}$, with each dataset represented as $D_i=(\mathcal{X}_i, \mathcal{Y}_i)$, where $\mathcal{X}_i$ refers to the time series of input signals and $\mathcal{Y}_i$ denotes the corresponding task outputs.

The signal $\mathcal{X}_i$ comprises sensor-monitored signals with variable dimensions $M_i$, distinct frequencies $f_i$, diverse operational conditions, and different outputs of tasks $\mathcal{Y}_i$. This variability leads to significant heterogeneity among the datasets $D_i$, making it extremely challenging to construct a unified model capable of diagnosing BAS systems across multiple aircraft types. 
Existing approaches often struggle to effectively capture the variations between different aircraft types, primarily because they are typically designed to handle homogeneous data from a single aircraft type. The differences in sensor configurations, operational frequencies, and task characteristics across aircraft types prevent the development of a generalized model, leading to substantial limitations in cross-aircraft type diagnosis. 
\subsection{Self-Supervised Pretraining Approach}
\label{subsec2}

Unlike traditional models that use individual functions $\mathcal{Y}_i = F_i(\mathcal{X}_i; \theta_i)$ for specific datasets $D_i$, our goal is to develop a foundation model $Y = F(\mathcal{X}; \theta)$ that works across all BAS datasets $D$. A promising approach is leveraging the capabilities of self-supervised learning, which excels at learning robust representations from unlabeled data and unifying diverse tasks and conditions.

In self-supervised learning, a pretext task $T_p$ is designed to train the model using unlabeled data $\mathcal{X}$. The objective is to learn a general representation $\mathbf{h} = F(\mathcal{X}; \theta_e)$, where $\theta_e$ are the pretraining parameters. This representation $\mathbf{h}$ is then used to solve the pretext task $T_p$, typically by predicting a part of the input $\mathcal{X}$ based on the rest. Formally, the self-supervised learning objective is:
\begin{equation}
\theta_e = \arg \min_{\theta_e} \mathbb{E}_{\mathcal{X} \sim D} \left[ \mathcal{L}_{T_p}(\mathcal{X}, F(\mathcal{X}; \theta_e)) \right],
\end{equation}
where $\mathcal{L}_{T_p}$ is the loss function associated with the pretext task $T_p$. Once pretrained, the model $F$ and representation $\mathbf{h}$ can be fine-tuned on downstream diagnosis and prognosis tasks using labeled data $\mathcal{Y}$. Pretraining on extensive unlabeled data enables the model to learn generalized features, making it adaptable to various diagnosis and prognosis tasks, thereby providing a comprehensive and flexible framework for BAS diagnosis.

For fine-tuning, the model parameters $\theta_f$ are adjusted based on labeled data:
\begin{equation}
\theta_f = \arg \min_{\theta_f} \mathbb{E}_{(\mathcal{X}_i, \mathcal{Y}_i) \sim D} \left[ \mathcal{L}(\mathcal{Y}_i, F(\mathcal{X}_i; \theta_e, \theta_f)) \right],
\end{equation}
where the fine-tuned parameters $\theta_f$ are shared across all datasets and tasks. Notably, the number of parameters to be optimized during fine-tuning, $\theta_f$, is significantly smaller than the number of parameters in the pretraining model, $\theta_e$, making the fine-tuning process more efficient.

\subsection{Downstream Tasks}
\label{subsec3}

Building on the general representation $\mathbf{h}$ obtained through the foundation model $F$, we define the relationship for task outputs as:
\begin{equation}
\hat{\mathcal{Y}}_i = \arg \max_{\mathcal{Y}_i} P(\mathcal{Y}_i \mid \mathbf{h}_i, \theta),
\end{equation}
where $\hat{\mathcal{Y}}_i$ is the predicted output for dataset $D_i$, based on the learned representation $\mathbf{h}_i$ and the foundation model $F$ with parameters $\theta_f$.

\subsubsection{Baseline Signal Prediction}
Baseline signal prediction (BP) involves predicting key monitored signals, such as gas pressure supplied by the BAS to downstream systems. The predicted baseline $\hat{\mathcal{Y}}^{BP}_i \in \mathbb{R}^{M_{Ba} \times T}$ provides the expected values of these key signals under normal operating conditions. This baseline serves as a reference for diagnostics, allowing for comparisons between actual and baseline signals. By comparing the true signals with the predicted baseline, we can determine the direction and extent of any deviations, providing interpretability in the diagnostic process.

\subsubsection{Anomaly Detection}

Anomaly detection (AD) involves identifying abnormal system behavior, represented by an anomaly label $\hat{\mathcal{Y}}^{AD}_i \in \mathbb{R}^{M_{An} \times T}$. The label takes a value of 0 when the system is operating normally and 1 when an anomaly is detected. This task aims to identify potential faults or irregularities, ensuring timely intervention and maintenance. Anomaly detection plays a crucial role in maintaining the reliability and safety of the BAS, as it enables early identification of issues before they lead to system failure.

\section{Methodology}

\begin{figure*}[hbtp]
    \centering
    \includegraphics[width = 1\textwidth]{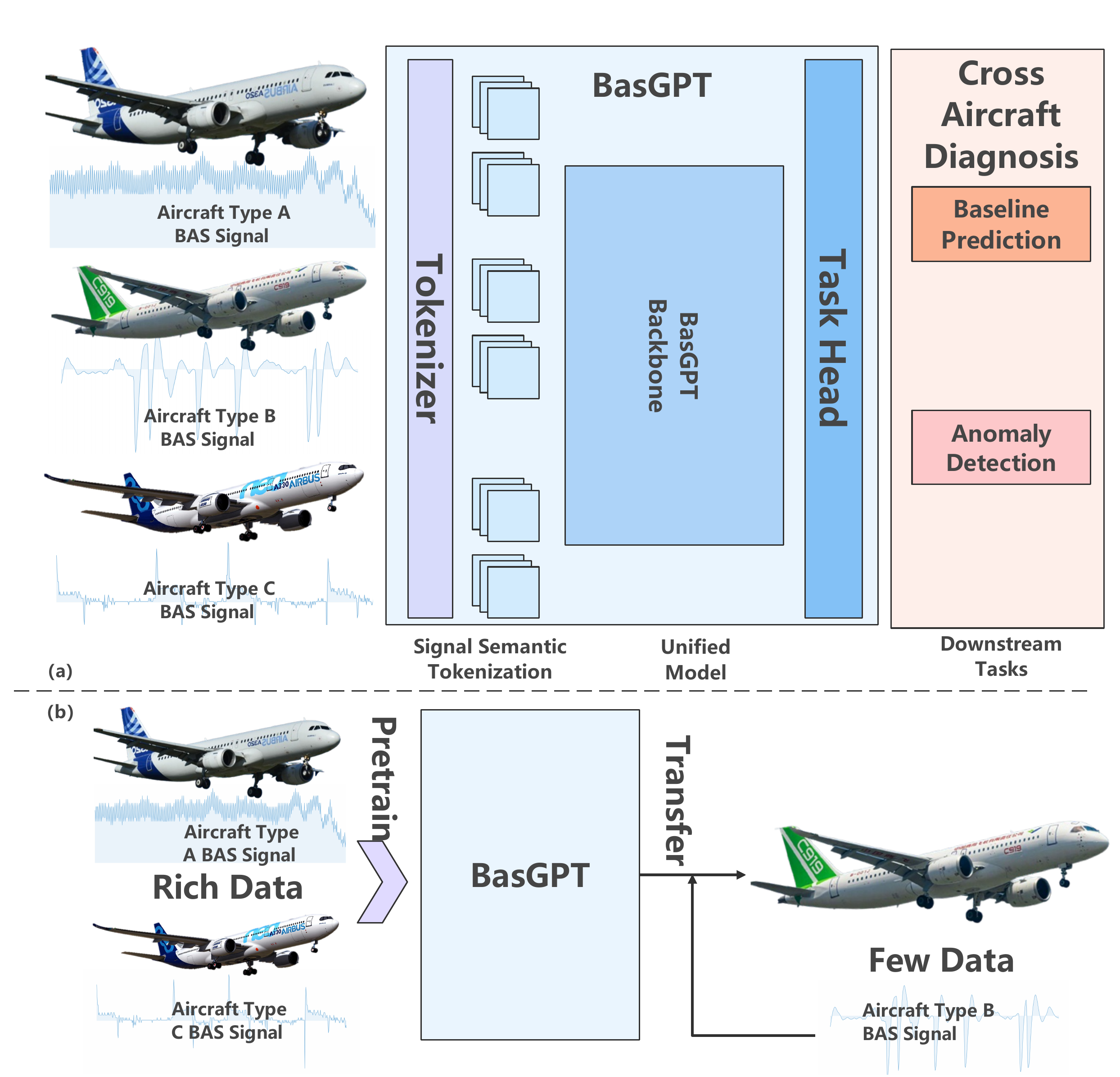}
    \caption{(a) The BAS Generative Pretrained Transformer (BasGPT) models the BAS system signals of three different aircraft types using a unified model architecture. (b) BasGPT is pretrained on BAS signals from mature aircraft types (e.g., A320, A330) and transfers the learned knowledge to a data-scarce new aircraft type (C919).}
    \label{fig:main}
\end{figure*}

In this chapter, we introduce the BAS Generative Pretrained Transformer (BasGPT), as shown in Fig. \ref{fig:main}. BasGPT serves as a foundation model for unified diagnosis and prognosis of BAS across different commercial aircraft. The model employs a generative framework that uses tokens to represent diagnostic tasks. It features an optimized architecture with a cross-aircraft type tokenizer and a dual-stage attention transformer capable of handling signals of varying channel counts and lengths, ensuring efficient processing and adaptability. The training process includes a pretraining strategy focused on reconstructing multi-dimensional masked signals, followed by fine-tuning for specific tasks.

Building a cross-aircraft type general diagnostic model, as depicted in Fig. \ref{fig:main}.(a), offers several benefits:

1.\textbf{ Enhanced Knowledge Sharing}: Although different aircraft have distinct BAS sensor groups due to varying designs, the underlying dynamics of the sensor signals are driven by shared physical principles, such as the second law of thermodynamics. By using a unified model, BasGPT can effectively extract and utilize this shared knowledge, improving the monitoring and diagnostic performance of BAS systems.

2. \textbf{Improved Knowledge Transfer}: When a new aircraft type begins operation, it often lacks sufficient operational data for data-driven monitoring. BasGPT addresses this challenge by leveraging knowledge gained from abundant historical data of maturely operated aircraft types, enabling effective knowledge transfer. This enhances the diagnostic capabilities and safety of the new aircraft during its early operational stages, as illustrated in Fig. \ref{fig:main}.(b).

 \subsection{Unified Model: BasGPT for Bleed Air System Diagnostics}

To develop a foundation model $\mathcal{F}$ capable of handling any input system $\mathcal{X}_i$, we must address two key challenges: (1) Differentiating and recognizing signals from various sensors, each with different physical properties, sampling rates, and magnitudes, while consistently extracting their internal semantics. (2) Efficiently modeling system signals $\mathcal{X}_i$ that have varying numbers of sensors ($M_i$) and different input lengths ($L_i$), using a backbone network to understand the relationship between signal semantics and diagnosis.

\subsubsection{Cross-Aircraft Type Flight Signal Tokenizer}

\begin{figure*}[hbtp]
    \centering
    \includegraphics[width = 1\textwidth]{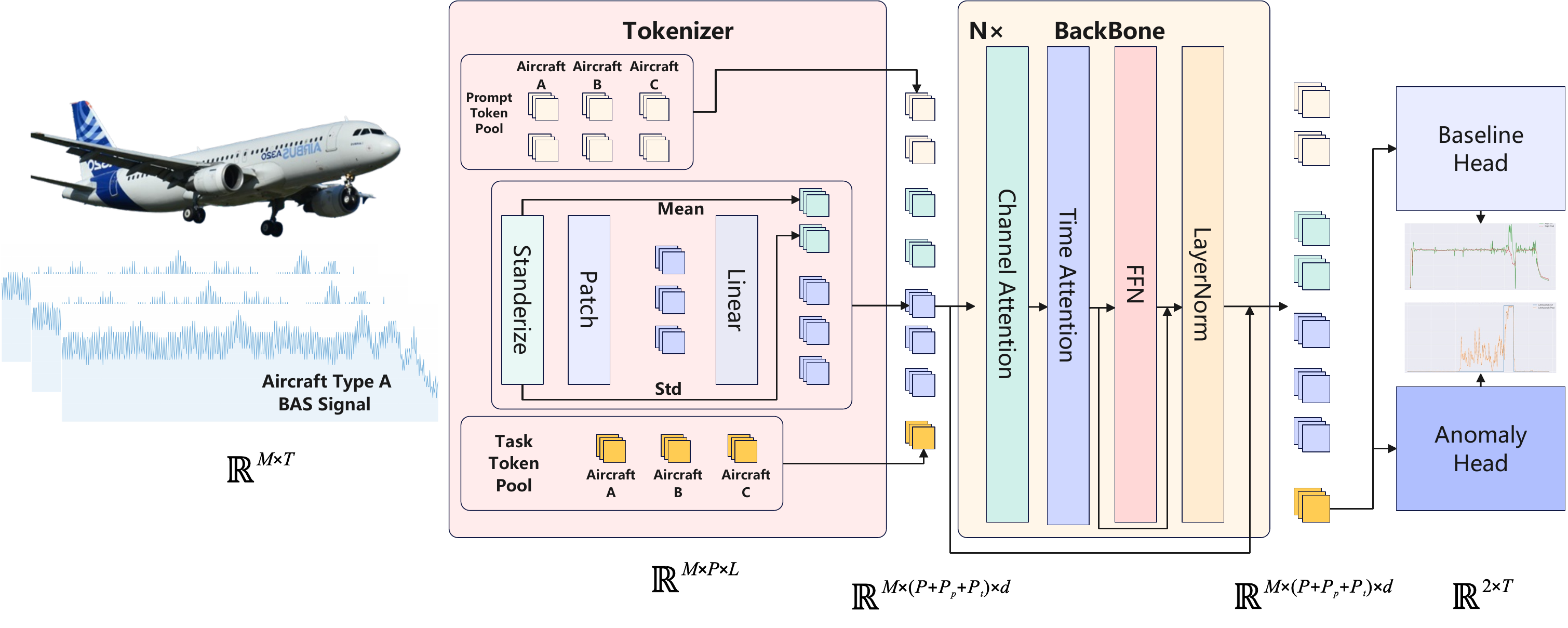}
    \caption{BasGPT model design: A Prompt token pool and CLS token pool for different aircraft types are prepended and appended to signal tokens in the Tokenizer to model inter-aircraft signal differences. An Efficient Self-Adaptive Transformer handles tokens of arbitrary length and channels to obtain high-level semantic representations, which are used for anomaly detection and baseline prediction.}
    \label{fig:framework}
\end{figure*}

To address the first challenge, we introduce the Cross-Aircraft Type Flight Signal Tokenizer (CFSTokenizer), designed to process input signals $\mathcal{X}_i$ from any aircraft type. CFSTokenizer, in Fig. \ref{fig:framework} can automatically distinguish between sensors from different aircraft types during training while consistently extracting semantic information for each sensor. Specifically, for an iput signal $\mathcal{X}_i$ of length $L_i$, we first normalize the signal to ensure stability. Each channel of the input signal is then divided into $P_i$ non-overlapping patches, each containing $pl$ points, spaced by $S$ points between patches. The transformation is given by:
\begin{equation}
P_i = \left\lfloor\frac{L_i - pl}{S}\right\rfloor + 1
\end{equation}\label{math:patch}
Each patch is then linearly projected into a latent space of dimension $d$. The normalized mean and variance are subsequently appended to each projected patch, resulting in the Signal Token $\mathcal{Z}_i^s \in \mathbb{R}^{M_i \times (P_i + 2) \times d}$, where $d$ denotes the size of the model's hidden layer.

Additionally, each sensor has two learnable tokens: a Prompt Token $\mathcal{Z}^p_{m,i} \in \mathbb{R}^{P_p \times d}$ and a Task Token $\mathcal{Z}^t_{i}\in \mathbb{R}^{1 \times P_t \times d}$. The Prompt Token encodes specific characteristics of the sensor, such as physical meaning and sampling rate, while the Task Token represents the characteristics of the aircraft type. These tokens are appended to the sequence of signal tokens to guide the model in recognizing the type of sensor and aircraft. The final token sequence input to the backbone network is represented as $\mathcal{Z}_i \in \mathbb{R}^{M_i \times (P_l + P + 2 + P_t) \times d}$, where:

\begin{align}
\mathcal{Z}_i &= \operatorname{CFSTokenizer}(\mathcal{X}_i) \\
              &= \operatorname{Concat}(\mathcal{Z}^p_i, \mathcal{Z}_i^s, \mathcal{Z}^t_i) \notag
\end{align}

\subsubsection{Efficient Self-Adaptive Transformer}

The Efficient Self-Adaptive Transformer (ESAT) aims to address two main challenges: (1) adapting to variable-length inputs to effectively extract features, and (2) reducing the computational burden typical of traditional Transformer networks, which scales quadratically with the number of tokens $N$ and channels $M$. To tackle this, we implement a two-stage attention mechanism that brings down the computational complexity from $O(N^2 \times M^2)$ to $O(N^2) + O(M^2)$, ensuring a more efficient process.

The first step involves channel attention, where tokens from different channels within the same time block are aggregated to derive the relevant semantics:
\begin{equation}
\operatorname{ChannelAttention}(\mathcal{Q}_c, \mathcal{K}_c, \mathcal{V}_c) = \operatorname{softmax}\left(\frac{\mathcal{Q}_c \mathcal{K}_c^T}{\sqrt{d}}\right) \mathcal{V}_c
\end{equation}
where $\mathcal{Q}_c$, $\mathcal{K}_c$, and $\mathcal{V}_c$ represent the query, key, and value matrices for the channel dimension.

The next step is time attention, which aggregates tokens across time segments within the same channel to derive the overall semantic information of the signal:
\begin{equation}
\operatorname{TimeAttention}(\mathcal{Q}_t, \mathcal{K}_t, \mathcal{V}_t) = \operatorname{softmax}\left(\frac{\mathcal{Q}_t \mathcal{K}_t^T}{\sqrt{d}}\right) \mathcal{V}_t
\end{equation}
where $\mathcal{Q}_t$, $\mathcal{K}_t$, and $\mathcal{V}_t$ are the query, key, and value matrices for the time dimension. Each ESAT block includes both attention mechanisms, followed by a Feed Forward Network (FFN) and a Layer Normalization (LayerNorm) layer. The backbone of the BasGPT model is constructed by stacking $N$ ESAT blocks in Fig. \ref{fig:framework}.

By employing these two attention mechanisms, the Efficient Self-Adaptive Transformer can effectively handle inputs of varying lengths and numbers of channels, ensuring computational efficiency without sacrificing essential information. This design significantly reduces the computational load, enhances feature extraction, and ultimately results in better diagnostic and prognostic performance.

 \subsection{Pretraining and Fine-Tuning Strategy for Flight Signal}

\subsubsection{Multidimensional Signal Token Masking}

Flight signal data is typically high-dimensional, consisting of various sensor channels with different sampling rates and time resolutions. To address this complexity, we introduce the Multidimensional Signal Token Masking (MSTM) task as part of our pretraining strategy in Fig.\ref{fig:pretrain}. This task is designed to learn the structure of flight signals by selectively masking parts of the signal and requiring the model to predict the missing portions based on the surrounding context.

Given an input flight signal $\mathcal{X}_i$, the signal is first divided into $P_i$ non-overlapping patches, as explained in the CFSTokenizer. After projecting the signal patches into a latent space, a subset of these tokens is randomly masked. Specifically, the masking process involves selecting a subset of patches \( Mask \subseteq \{1, 2, ..., P_i\} \) and replacing the values of the masked patches with the corresponding Task Token $\mathcal{Z}_i^t$ of the aircraft type. This forces the model to gather global information from the Task Token and predict the missing signal tokens using the unmasked patches.

The masked signal token sequence, denoted as $\hat{\mathcal{Z}}_i^s$, is computed as:

\begin{equation}
    \hat{\mathcal{Z}}_i^s = \mathcal{Z}_i^s \odot Mask + \mathcal{Z}_i^t \odot (1 - Mask)
\end{equation}

where \( Mask \) is a binary mask matrix, with \( Mask_j = 0 \) for masked patches and \( Mask_j = 1 \) for unmasked patches. The \( \odot \) operator denotes element-wise multiplication. During pretraining, the model learns to predict the masked patches using the unmasked tokens, guided by the global Task Token.

The prediction of the masked signal $\hat{\mathcal{X}}_i'$ is then compared to the original signal $\mathcal{X}_i$ using a loss function, such as Mean Squared Error (MSE), to evaluate the accuracy of reconstruction:

\begin{equation}
    \mathcal{L}_{\text{mask}} = \frac{1}{|Mask|} \sum_{i \in Mask} \left( \mathcal{X}_i - \hat{\mathcal{X}}_i' \right)^2
\end{equation}

This task encourages the model to learn both spatial and temporal dependencies within the signal patches and captures the underlying semantic patterns, such as flight and health status, through self-supervised learning. These learned representations are crucial for downstream tasks like fault detection and system diagnosis.

\begin{figure}[!htbp]
\centerline{\includegraphics[width=0.98\columnwidth]{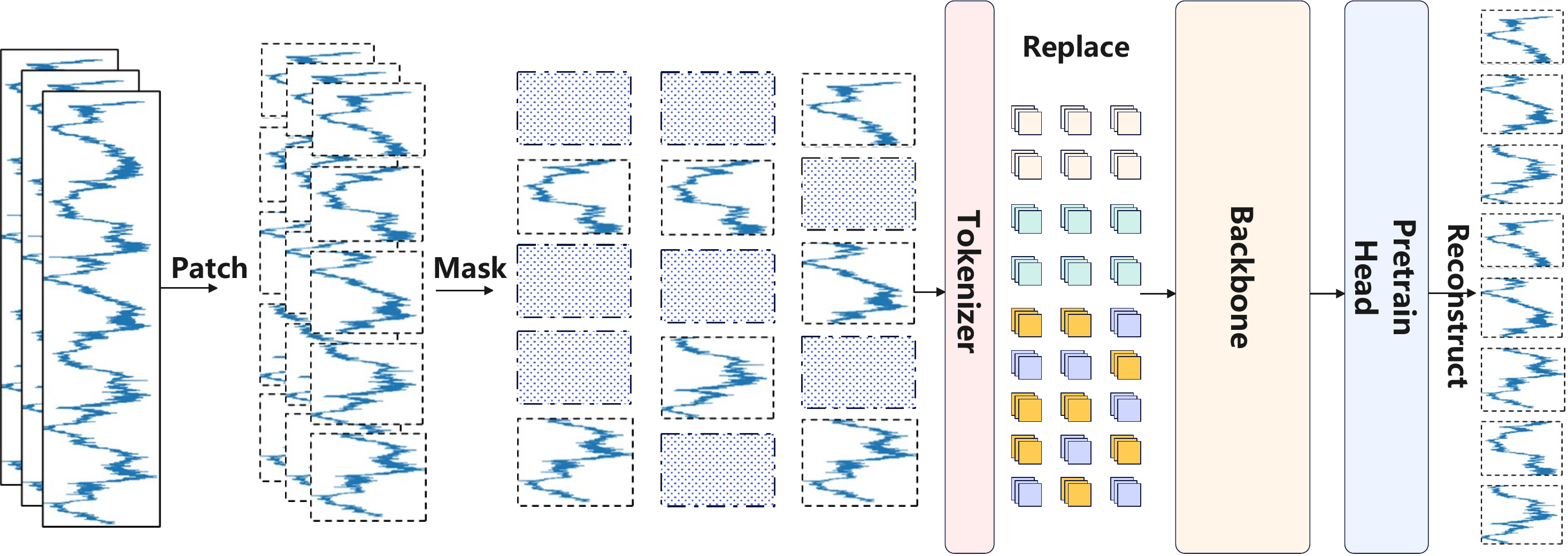}}
\caption{Multidimensional Signal Patch Masking Pretraining}
\label{fig:tokens and attention}
\end{figure}

 \subsubsection{Joint Baseline Prediction and Anomaly Detection Loss Function}

The model's final objective involves two primary downstream tasks: Baseline Prediction (BP) and Anomaly Detection (AD). BP aims to model the normal dynamics of an aircraft's Bleed Air System (BAS) signals by training on sensor data under normal operating conditions. This approach establishes the relationships between sensor parameters and target parameters during regular operations. Such modeling is widely applied in aero-engine gas path fault diagnosis to determine whether the engine is operating normally and to promptly analyze and isolate potential faults\cite{wang2024data}.AD focuses on identifying deviations from normal system states by defining the normal state as 0 and the fault state as 1, and training on these fault labels to achieve the goal of fault detection. AD methods\cite{Jiang2021Ano} are extensively used in industrial equipment anomaly detection by monitoring the state signals of equipment to promptly detect and distinguish abnormal events from normal ones.  While both tasks aim to understand the system's typical behavior, they are fundamentally different: BP focuses on accurate signal predictions under normal conditions, whereas AD identifies abnormal patterns.

A natural approach is to jointly train the model on both tasks, allowing it to learn the relationship between the normal state signals $\mathcal{X}_i$ and their corresponding BP predictions $\mathcal{Y}_i^{BP}$, while avoiding fitting the model to abnormal signals. This leads to the BP task loss function:

\begin{equation}
    \mathcal{L}^{BP} = (1 - \mathcal{Y}_i^{AD}) \odot \frac{1}{L_i} \sum_{i \in L_i} \left( \mathcal{Y}_i^{BP} - \hat{\mathcal{Y}}_i^{BP} \right)^2
\end{equation}

Here, when the system is in its normal state ($\mathcal{Y}_i^{AD} = 1$), the signal samples contribute to the model's gradient and are used for optimization. This ensures that the BP loss is calculated only for normal states.

For anomaly detection, the objective is to detect and classify anomalies at a finer level, providing a loss function that measures the prediction error at each time step. We use the MSE between the predicted anomaly label $\hat{\mathcal{Y}}_i^{AD}$ and the true anomaly label $\mathcal{Y}_i^{AD}$:

\begin{equation}
    \mathcal{L}^{AD} = \frac{1}{L_i} \sum_{i \in L_i} \left( \mathcal{Y}_i^{AD} - \hat{\mathcal{Y}}_i^{AD} \right)^2
\end{equation}

Finally, the total loss for joint training is the weighted sum of the BP and AD losses:

\begin{equation}
    \mathcal{L}^{total} = \mathcal{L}^{BP} + \alpha \times \mathcal{L}^{AD}
\end{equation}

where $\alpha$ is a hyperparameter that controls the relative importance of the BP and AD losses. Since most flight signals are normal, $\alpha$ is typically set to a value greater than 5 to ensure the model prioritizes normal signal prediction.

During fine-tuning, the model is efficiently adapted to different aircraft types by training only a small subset of parameters. Specifically, we fine-tune only the two projection layers $\theta_f$, which constitute less than 1‰ of the total model parameters. The parameters of the core model, including the CFSTokenizer and Backbone network, are frozen and remain fixed as $\theta_e$. This allows the model to efficiently generalize across various aircraft systems while retaining its prelearned knowledge.

\section{Experimental Results}
In this section, we present a comprehensive evaluation of the BasGPT model across various baseline prediction and anomaly detection tasks, leveraging a range of diverse aircraft datasets. We constructed A320, A330, C919 datasets based on industrial real flight data through data collection, data cleaning, data filtering, feature selection, normalization, and other operations, providing a factual basis for this study. Next, we report the main experimental results, comparing the performance of BasGPT on two downstream tasks with state-of-the-art algorithms \cite{wutimesnet,liu2022pyraformer,liuitransformer,zeng2023transformers}. In addition, we pretrain the model on datasets from legacy aircraft systems and fine-tune it on a set of newer aircraft data, demonstrating the significant benefits of transfer learning from mature aircraft types to new aircraft platforms. Also, we analyzed the relationship between model performance and model capacity, providing a foundation for future large-scale model development and demonstrating the growth value of model scale in real industrial scenarios. Finally, we provided the visualization analysis of learned representations with different aircraft types using t-SNE.

\subsection{Dataset Description}

In this study, we have curated and organized BAS datasets for A320, A330, and C919 aircraft to evaluate the performance of the BasGPT model in baseline prediction and anomaly detection tasks. Table \ref{tab:dataset} provides a summary of these datasets, which come from different aircraft types, including signal channels, fault categories, sampling frequency, and dataset size.
\begin{table}[htbp]
\caption{Overview of datasets}
\label{tab:dataset}
\centering
\setlength{\tabcolsep}{1.5pt} 
\renewcommand{\arraystretch}{1.1} 
\begin{tabular}{@{}p{1.8cm}p{1.8cm}p{4.0cm}p{1.8cm}p{2cm}@{}}
\toprule
 Datasets & Channels & Fault & Total
 Points& Sample Frequency \\ \midrule
 A320 & 52 & under-pressure & 936512 & 0.2Hz \\
 A330 & 37 & over-pressure & 314132 & 1Hz\\
 C919 & 17 & over-temperature & 445953 & 0.25Hz \\ \midrule
\end{tabular}
\end{table}

The datasets mentioned above were collected from the latest flight data, which have real-world industrial scenarios and are problems encountered during actual operation. The collection of BAS system sensor data follows the airworthiness standards outlined in \cite{CAAC2017}, ensuring compliance with regulatory requirements for transport category aircraft. This dataset, for the first time, has been systematically collected and studied in this research. However, due to significant variations between these datasets, including differences in data size and the number of faults, they present unique challenges. Previous studies have not attempted to apply models with the same parameters for both baseline prediction and anomaly detection tasks across datasets from different aircraft BAS systems. This work aims to demonstrate the feasibility of using a single model to effectively handle data from multiple aircraft types and achieve good performance in downstream tasks. Furthermore, we show that this universal modeling capability allows the model to leverage diagnostic knowledge from mature operation aircraft types (such as A320 and A320), transfer this knowledge to newer aircraft types, and achieve notable performance improvements.

\subsection{Main Experimental Results}
\subsubsection{Experimental Setup}
In our experiment, suitable hyperparameters were obtained through multiple rounds of experimental optimization to train the BasGPT model, ensuring optimal performance in various diagnostic and prognostic tasks. The experiments are conducted on a computing server equipped with 8 NVIDIA RTX 3090 GPUs. 

The detailed hyperparameters are summarized in Table \ref{tab:hyperparameters}. To minimize differences in the input space, standardize the data and select cruise level flight phase for our research. Each input signal window is standardized to 2048 steps. In the Tokenizer settings, the stride length and patch length are both set to 128. In order to ensure that the model has sufficient representation ability, we conducted multiple experiments at different layers and model depths. The specific results will be presented in Model Scalability Analysis. Finally, we chose to configure the model architecture as 4 layers and 512 hidden units, with a total of 24.61M parameters, which has high efficiency while ensuring the model's ability. For the length of prompts and fault tokens, we use lengths of 10 and 1 respectively. In the optimization phase, the model is trained with a batch size of 256 and the learning rate is set, respectively. During the optimization phase, the model is trained with a batch size of 256 and the learning rate is set to \(3.00 \times 10^{-7}\). The pretraining phase consists of 5 epochs, while the fine-tuning phase consist of 5 epoch.

\begin{table}[htbp]
\caption{Hyperparameters Used for Training BasGPT Model}
\label{tab:hyperparameters}
\centering
\setlength{\tabcolsep}{2pt} 
\renewcommand{\arraystretch}{1.2} 
\begin{tabular}{@{}cc@{}}
\toprule
\multicolumn{2}{c}{Hyperparameters} \\ \midrule
Batch Size                & 256       \\
Learning Rate             & \(3.00 \times 10^{-7}\)  \\
Pretrain Epochs           & 5        \\
Finetune Epochs           & 5         \\
Stride Length ($S$)         & 128       \\
Patch Length ($P$)          & 128       \\
Layers                    & 4         \\
Hidden Numbers ($d$)        & 512       \\
Prompt Token Length ($l_p$)& 10        \\
Fault Token Length ($l_t$) & 1         \\ 
Model Size                 &24.61 M     \\\bottomrule
\end{tabular}
\end{table}

\subsubsection{Key Results}
We present the key experimental results, showcasing the performance of the BasGPT model on baseline prediction and anomaly detection tasks across multiple datasets. The results, summarized in Table \ref{tab:baseline}, highlight the effectiveness of the model in baseline prediction. In the table, “BasGPT (Pretrained)” refers to the results obtained after pretraining on unlabeled data from three aircraft types followed by fine-tuning, while “BasGPT” represents the results from direct fine-tuning without pretraining.

For the three different datasets, we utilized a single BasGPT model for both training and testing, whereas the SOTA methods were trained separately for each dataset. The results indicate that our joint training approach yields more accurate predictions compared to traditional individual training methods.

BP, the first downstream task, assesses the model's ability to reconstruct expected system behavior under normal operating conditions. On the C919 dataset, which contains limited data, BasGPT demonstrates a clear advantage in BP performance. This improvement is attributed to the pre-training phase, where the model learns generic feature representations from diagnostic knowledge of different aircraft types, enabling effective knowledge transfer to data-scarce scenarios. As a result, BasGPT achieves significantly lower Mean Absolute Error (MAE) and Mean Squared Error (MSE) compared to conventional methods, with an MAE of 0.0606 and an MSE of 0.0209. These results, presented in Table~\ref{tab:baseline}, highlight the model's superior ability to generalize across different aircraft types and enhance diagnostic accuracy even in data-limited settings.

\begin{table}[!htbp]
\caption{BasGPT Uses the Same Parameter Model and Outperforms SOTAs Algorithms on baseline Prediction}
\label{tab:baseline}
\centering
\setlength{\tabcolsep}{12pt} 
\renewcommand{\arraystretch}{1.2} 
\resizebox{\textwidth}{!}{ 
\begin{tabular}{@{}ccccccc@{}}
\toprule                     
Dataset         &\multicolumn{2}{c}{A320}               & \multicolumn{2}{c}{A330}        & \multicolumn{2}{c}{C919}        \\
Metric          & MAE↓           & MSE↓         & MAE↓           & MSE↓& MAE↓           & MSE↓                                      \\ \midrule
BasGPT (Pretrained)  & \textbf{0.2308}& \textbf{0.1931}  & \textbf{0.2101}& \underline{0.3194}& \textbf{0.0606}& \textbf{0.0209}    \\
BasGPT & \underline{0.2314}& \underline{0.1976} & 0.2220& 0.3363& \underline{0.0632}& \underline{0.0220} \\
TimesNet\cite{wutimesnet}   &0.2752 &  0.2173   & 0.3012& 0.5124& 0.1217&0.0512        \\
Pyraformer\cite{liu2022pyraformer}       &0.2752 &  0.2173     & 0.2669& 0.3713& 0.0943& 0.0483        \\
iTransformer\cite{liuitransformer}   &0.2752 &  0.2173     & 0.2636& 0.4627& 0.0812& 0.0318         \\
Dlinear\cite{zeng2023transformers}   &0.2752 &  0.2173      & \underline{0.2110}& \textbf{0.3017} & 0.0649& 0.0299        \\
\end{tabular}}
\end{table}

The second downstream task that we evaluate is AD. For each model in the test set, we recorded the number of True Positives (TP), False Positives, True Negatives, and False Negatives, where fault cases are considered as positive examples. Based on these counts, we computed the Precision(Prec.), Recall(Rec.), and F1 scores(F1), which are summarized in Table \ref{tab:baseline}.

The results demonstrate the performance of anomaly detection across different datasets. Notably, we observe that on various datasets, BasGPT outperforms other methods in terms of both Precision and Recall. This trend is especially pronounced on datasets with limited fault instances, such as C919 and A320. Because lack of self-supervised pretraining, traditional deep learning models struggle to learn the features of fault data in such scenarios due to the absence of sufficient labeled examples. In fact, for these datasets, many of the traditional methods show values of 0 for TP or "None" for precision and F1-score, indicating that the model failed to identify any relevant fault cases.

\begin{table}[!htbp]
\caption{BasGPT Uses the Same Parameter Model and Outperforms SOTA Algorithms on Anomaly Detection}
\label{tab:baseline}
\centering
\setlength{\tabcolsep}{4pt} 
\renewcommand{\arraystretch}{1.5} 
\resizebox{\textwidth}{!}{ 
\begin{tabular}{@{}lccc|ccc|ccc@{}}
\toprule 
Method  & \multicolumn{3}{c|}{A320}   & \multicolumn{3}{c|}{A330}   & \multicolumn{3}{c}{C919}        \\
        & Prec.↑   & Rec.↑  & F1↑   & Prec.↑   & Rec.↑  & F1↑   & Prec.↑    & Rec.↑  & F1↑      \\ \midrule
BasGPT (Pretrained)  & \textbf{0.878} & \textbf{0.912} & \textbf{0.950} & 0.922 & \underline{0.591} & \underline{0.720} & \textbf{0.949} & \textbf{0.905} & \textbf{0.927}  \\
BasGPT  & None & 0 & None & \underline{0.952} & 0.261 & 0.410 & None & 0 & None   \\
TimesNet\cite{wutimesnet}   & None & 0 & None  & 0.938 & 0.221 & 0.358 & None & 0 & None       \\
Pyraformer\cite{liu2022pyraformer}   & None & 0 & None & \textbf{0.998} & 0.308 & 0.470 & None & 0 & None     \\
iTransformer\cite{liuitransformer} & 0.112 & \underline{0.429} & 0.178 & 0.815 & \textbf{0.665} & \textbf{0.732} & None & 0 & None\\
Dlinear\cite{zeng2023transformers}   & \underline{0.170} & 0.357 & \underline{0.231} & 0.893 & 0.318 & 0.469 & 0 & 0 & None 
     \\
\end{tabular}}
\end{table}

\begin{figure}[htbp]
\centerline{\includegraphics[width=0.98\columnwidth]{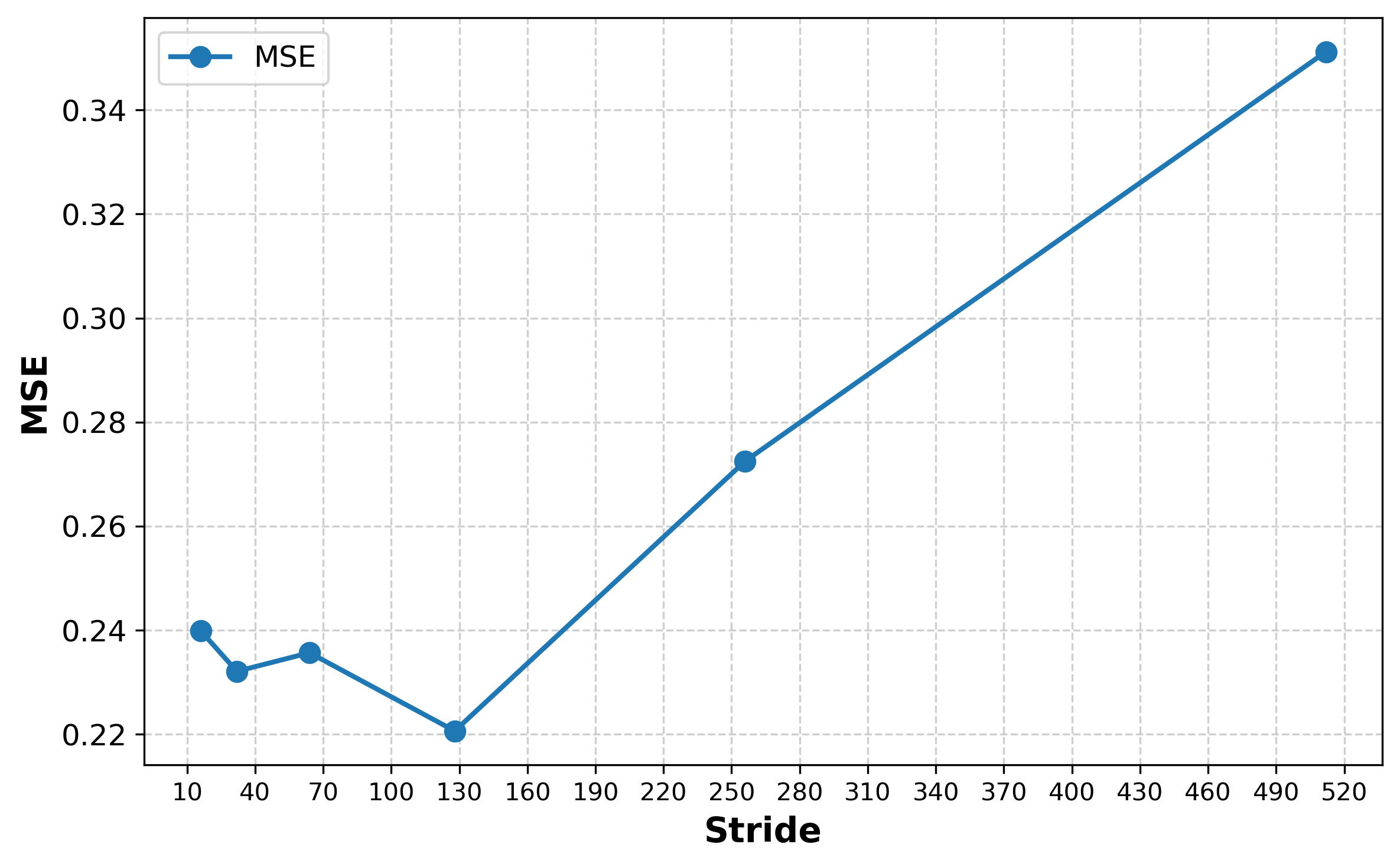}}
\caption{Performance of the model on the BP task across three aircraft types, varying with different stride and patch length.}
\label{fig:s}
\end{figure}

To investigate the impact of stride length ($S$) and patch length ($P$) on model performance, we conducted ablation experiments with $P = S$ to ensure non-overlapping dataset sampling. A key trade-off exists between computational efficiency and feature representation: smaller $S$ increases tokenized patches, capturing fine-grained local features but also introducing noise, while larger $S$ reduces computational cost but may lose critical details. As shown in Fig. \ref{fig:s}, model performance first improves and then declines with increasing $S$, achieving optimal results at $P = S = 128$. This suggests that a moderate $S$ balances local and global feature extraction while maintaining efficiency, underscoring the importance of carefully selecting $S$ and $P$.

These results underscore the importance of self-supervised pretraining. Through this approach, BasGPT is able to learn universal feature representations from flight signal data without requiring labeled fault data. And demonstrate the feasibility of using self-supervised method to diagnose and predict on different types of aircraft. And by using the Joint Baseline and Anomaly Detection Loss Function, enabling efficient fine-tuning and interpretable anomaly detection.

\subsection{Case Study}
\begin{figure}[!htbp]
\centerline{\includegraphics[width=0.98\columnwidth]{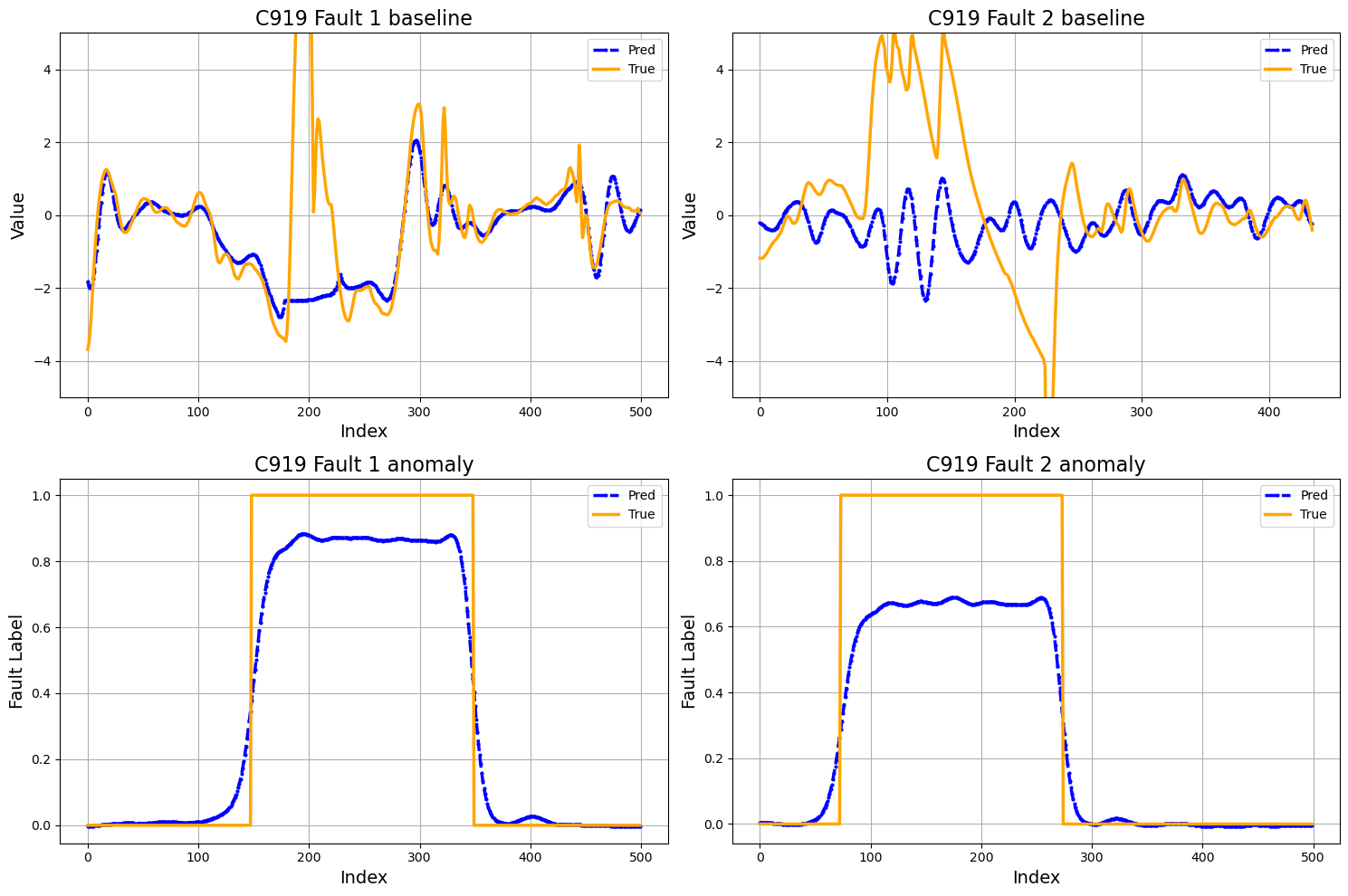}}
\caption{These are two C919 fault cases, two figures above are baseline prediction and below are anomaly detection.}
\label{fig:baseline}
\end{figure}

Here, we present some cases of baseline prediction and anomaly detection, and demonstrate their correlation. Because the baseline prediction and fault detection of the BAS of A330 have already been studied in \cite{wang2024data}, the results are mainly presented here on a small dataset C919.
In Fig. \ref{fig:baseline}, it shows the baseline prediction results for the fault segment and normal segment.

The top two figures in Fig. \ref{fig:baseline} show the actual temperature values and the baseline predicted values during two fault periods of C919 aircraft. The following two figures illustrate the anomaly detection results for the same time intervals. During these abnormal periods, the actual temperature values significantly deviate from the baseline predictions, highlighting a clear anomaly. The baseline prediction, which models the expected normal behavior of the aircraft's BAS signals, reflects the system's typical operational dynamics. In contrast, anomaly detection focuses on identifying deviations from this baseline, which in these cases corresponds to faults in the BAS system. The significant difference between the actual and predicted values during the fault periods supports the reliability of the anomaly detection results, providing an interpretable diagnosis of the aircraft’s BAS performance.

This pattern of performance has been consistently observed across other models as well. However, due to space constraints, these results are not presented in this figure.

\subsection{Analysis of Transferability}
To validate the proposed model's ability to extract universal feature representations from flight signals without relying on labeled data, achieved through self-supervised pre-training, we conducted experiments to evaluate its transfer learning performance. The experiments compared the transfer learning performance of our method (pre-trained on source datasets) against a baseline (no pre-training). We measured performance on target datasets in terms of MAE and MSE, as shown in Table \ref{tab:transfer}.

\begin{table}[htbp]
\caption{Transferability Analysis}
\label{tab:transfer}
\centering
\setlength{\tabcolsep}{1.5pt} 
\renewcommand{\arraystretch}{1.1} 
\begin{tabular}{@{}p{4cm}p{2cm}p{2.0cm}p{1.2cm}p{1.2cm}@{}}
\toprule
 Source Datasets & Transfer Dataset & MAE↓ & MSE↓ \\ \midrule
 A330, C919 & A320 & \textbf{0.2310} & \textbf{0.1946}  \\
 None& A320 & 0.2514 & 0.2076 \\
 A320, C919 & A330 & \textbf{0.2215} & \textbf{0.2284}  \\ 
 None& A330 & 0.2513 & 0.3934 \\ 
 A330, A320  & C919 & \textbf{0.0613} & \textbf{0.0211} \\
 None&  C919 & 0.0832 & 0.0320 \\ \midrule
\end{tabular}
\end{table}

 For example, when transferring to C919 dataset, a new aircraft type with limited labeled fault data, the model achieved a notable improvement with an MAE of 0.0613 and an MSE of 0.0211, compared to 0.0832 and 0.0320 for the baseline. Similarly, for A320 dataset, the proposed method reduced MAE and MSE from 0.2514 and 0.2076 (baseline) to 0.2310 and 0.1946, respectively. These results highlight the model’s ability to transfer diagnostic knowledge from mature aircraft, such as A320 and A320, to datasets with sparse fault data or new operational environments. This capability ensures robust fault detection and prediction, even in early-stage operations of new aircraft, where labeled fault data is scarce or unavailable.

\subsection{Model Scalability Analysis}

We conducted experiments to evaluate the impact of model parameter scaling on predictive performance, focusing on models pre-trained on A320 and A330 datasets and transferred to C919 dataset. Specifically, we varied the model depth (\(d\)) across values of 64, 128, 256, 512, 1024, and 2048, while testing different layer configurations (1, 2, 3, and 4 layers). Predictive performance was measured using the Mean BASolute Percentage Error (MAPE), as it normalizes the scale of error, avoiding distortion caused by high-amplitude time series data that can affect metrics like MAE and MSE.

\begin{figure}[!htbp]
\centerline{\includegraphics[width=0.98\columnwidth]{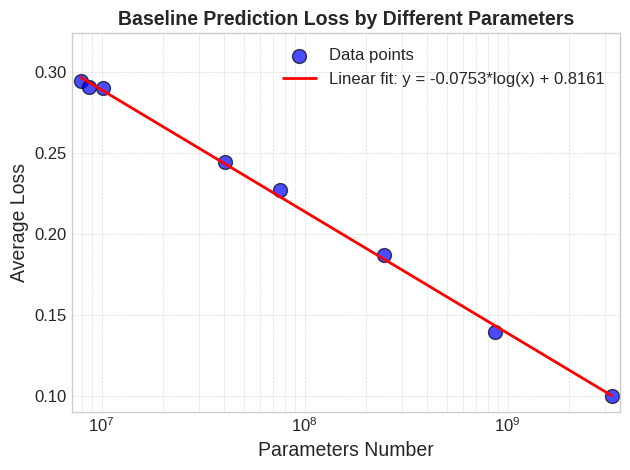}}
\caption{Parameter Scaling. The impact of model parameter scaling on forecasting performance, evaluated using MAPE.}
\label{fig:pretrain}
\end{figure}

In Fig. \ref{fig:pretrain}, we illustrate the performance of pre-trained models with parameter counts ranging from 0.79M (smaller models) to 325M (larger models). These experiments were conducted on C919 dataset after pre-training on A320 and A330 datasets, with MAPE reported as the primary performance metric. The observed results align with the power-law relationship described in \cite{yao2024towards}:

\begin{equation}
L(N) \approx\left(\frac{N_{c}}{N}\right)^{\alpha_{N}}
\end{equation}

where \emph{L} represents the performance metric (MAPE in this study), \emph{N} is the parameter count, \(N_{c}\) is the normalization coefficient, and \(\alpha_{N}\) is the exponent reflecting the rate of performance improvement with increased parameters.

The results in Fig. \ref{fig:pretrain} confirm that model performance improves consistently with parameter scaling, following the theoretical power-law relationship. This validates the feasibility of employing larger models in industrial scenarios, particularly for integrating diverse aircraft datasets into systems like the BAS and designing models tailored for other aircraft systems.

Moreover, this conclusion extends beyond aviation applications, highlighting that scaling up model parameters can achieve superior results in various industrial domains. This demonstrates the potential of leveraging larger models and datasets to address challenges in industrial data applications effectively.

\subsection{Feature Visualization}
\begin{figure}[!htbp]
\centerline{\includegraphics[width=0.98\columnwidth]{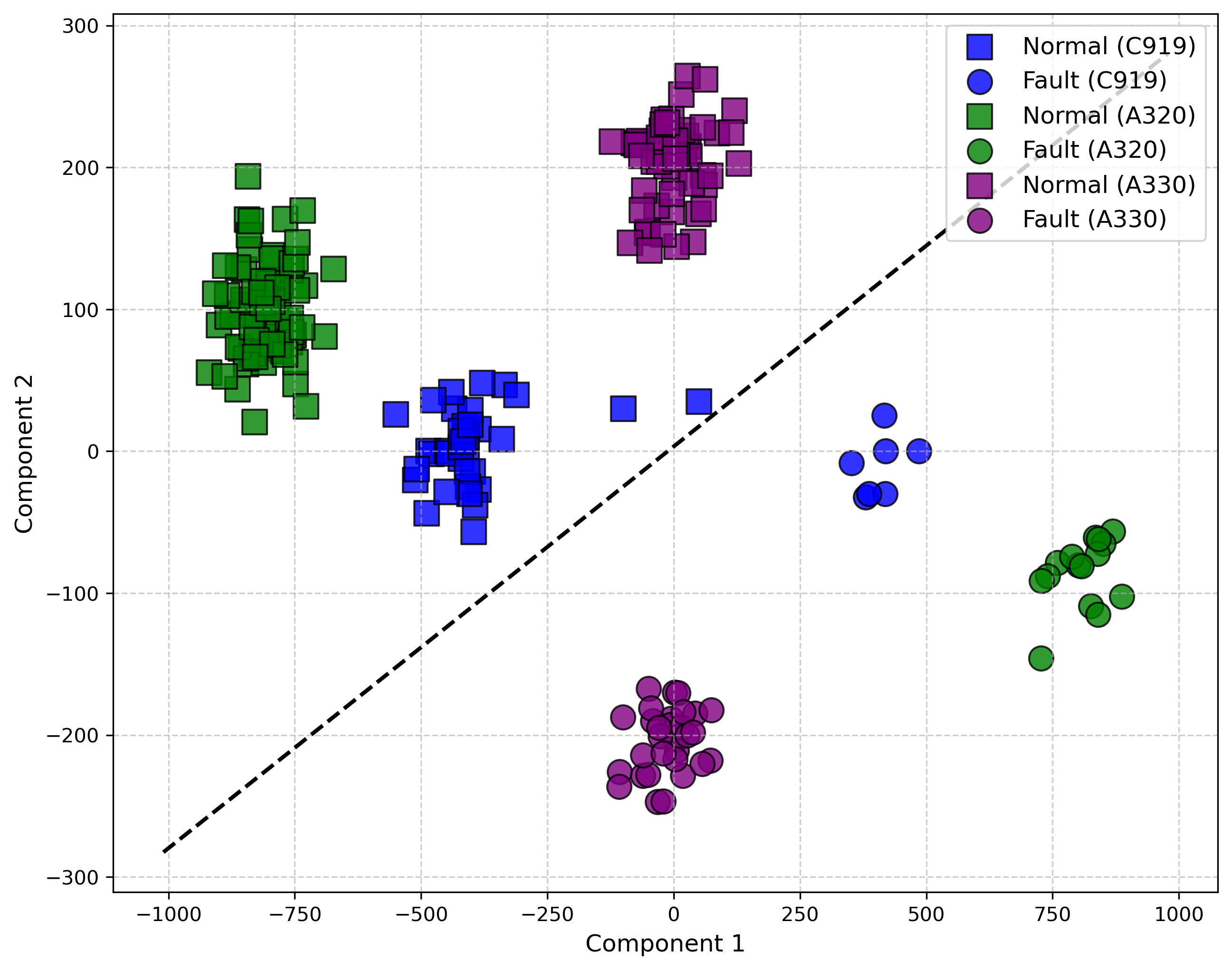}}
\caption{Visualization of fault and anomaly representations for different aircraft types using t-SNE, projected onto a 2D plane.}
\label{fig:feature}
\end{figure}

Figure \ref{fig:feature} illustrates the semantic representations of patch signals for different aircraft types under normal and anomalous conditions, as extracted by the model and visualized in a 2D space using t-SNE \cite{van2008visualizing}. Different aircraft types are distinguished by colors, with square markers representing normal states and circular markers indicating anomalies. Despite variations in semantic representations due to design and sensor differences among aircraft types, the model effectively delineates a clear boundary between normal and anomalous states in the learned representation space. This demonstrates that the BasGPT model successfully models BAS states across different aircraft types and accurately distinguishes normal from anomalous conditions.

The well-defined decision boundaries in the t-SNE visualization highlight BasGPT’s ability to learn shared semantic structures across aircraft types, enabling it to generalize effectively despite differences in sensor data distributions. This indicates that BasGPT captures essential fault-related features and can process heterogeneous sensor data while maintaining consistent fault classification. Its robustness in distinguishing normal and abnormal states across different aircraft types underscores its effectiveness in real-world diagnostics.

\section{Conclusion}
This paper introduces BasGPT, a self-supervised learning model designed for diagnosing and predicting faults in BAS across multiple aircraft types. The model excels in baseline prediction and anomaly detection, especially for new aircraft like C919 with limited fault data. By transferring knowledge from mature aircraft types, BasGPT addresses the challenge of sparse fault data and enhances diagnostic accuracy. Our experiments demonstrate that BasGPT outperforms traditional methods, achieving better precision and recall, particularly in datasets with limited fault instances.

The model scalability analysis shows that increasing the number of model parameters leads to improved performance, highlighting the potential for larger models to achieve even greater accuracy. This paves the way for developing larger-scale flight signal models in the future. Potential directions include incorporating data from other key aircraft components (e.g., Engine, Air Conditioner, Auxiliary Power Unit.etc) and building unified cross-device flight parameter models to create more intelligent and scalable systems.

In conclusion, BasGPT offers a powerful, adaptable solution for BAS diagnostics, with significant potential to enhance safety and operational efficiency across a range of aircraft types.

\section{Acknowledgements}
This work was partially supported by Grant of National Natural Science Foundation (Grant No. 62371297) and Fund of Shanghai Engineering Research Center of Civil Aircraft Health Monitoring (Grant No. GCZX202204).




\bibliographystyle{elsarticle-num}
\bibliography{main}
\end{document}